\documentclass[sigconf]{acmart}
\usepackage[german,american]{babel} 

\usepackage[T1]{fontenc}
\usepackage[latin1]{inputenc} 

\usepackage{xspace}

\newcommand{\Ni}{(1)~}
\newcommand{\Nii}{(2)~}
\newcommand{\Niii}{(3)~}
\newcommand{\Niv}{(4)~}
\newcommand{\Nv}{(5)~}

\newcommand{\bslabel}[1]{\textsl{#1.}}

\usepackage{booktabs} 

\usepackage{graphicx}
\DeclareGraphicsExtensions{.pdf,.ai,.jpg,.png}
\setkeys{Gin}{pagebox=artbox}
\graphicspath{{./}}

\newcommand{\bsfigure}[5][scale=1.0]{%
  \begin{figure}[tb]
    \centering
    \includegraphics[#1]{#2}
    \caption{#3}\label{#2}
    \Description[#4]{#5}
  \end{figure}}

\RequirePackage{type1cm}
\RequirePackage{xcolor}
\RequirePackage{soul}
\setstcolor{blue}

\setcopyright{iw3c2w3}

\copyrightyear{2020}
\acmYear{2020}
\acmConference[WWW '20]{Proceedings of The Web Conference 2020}{April 20--24, 2020}{Taipei, Taiwan}
\acmBooktitle{Proceedings of The Web Conference 2020 (WWW '20), April 20--24, 2020, Taipei, Taiwan}
\acmPrice{}
\acmDOI{10.1145/3366423.3380206}
\acmISBN{978-1-4503-7023-3/20/04}

\usepackage{silence}
\begin{document}

\title{Abstractive Snippet Generation}

\settopmatter{authorsperrow=5}

\author{Wei-Fan Chen}
\affiliation{%
\institution{Paderborn University}
}

\author{Shahbaz Syed}
\affiliation{%
\institution{Leipzig University}
}

\author{Benno Stein}
\affiliation{%
\institution{Bauhaus-Universit\"at Weimar}
}

\author{Matthias Hagen}
\affiliation{%
\institution{\mbox{\kern-0.8em Martin-Luther-Universit\"at} Halle-Wittenberg}
}

\author{Martin Potthast}
\affiliation{%
\institution{Leipzig University}
}

\begin{abstract}
An {\em abstractive snippet} is an originally created piece of text to summarize a web page on a search engine results page. Compared to the conventional {\em extractive snippets}, which are generated by extracting phrases and sentences verbatim from a web page, abstractive snippets circumvent copyright issues; even more interesting is the fact that they open the door for personalization. Abstractive snippets have been evaluated as equally powerful in terms of user acceptance and expressiveness---but the key question remains: Can abstractive snippets be automatically generated with sufficient quality?

This paper introduces a new approach to abstractive snippet generation: We identify the first two large-scale sources for distant supervision, namely anchor contexts and web directories. By mining the entire ClueWeb09 and ClueWeb12 for anchor contexts and by utilizing the DMOZ Open Directory Project, we compile the Webis Abstractive Snippet Corpus~2020, comprising more than 3.5~million triples of the form $\langle$query, snippet, document$\rangle$ as training examples, where the snippet is either an anchor context or a web directory description in lieu of a genuine query-biased abstractive snippet of the web document. We propose a bidirectional abstractive snippet generation model and assess the quality of both our corpus and the generated abstractive snippets with standard measures, crowdsourcing, and in comparison to the state of the art. The evaluation shows that our novel data sources along with the proposed model allow for producing usable query-biased abstractive snippets while minimizing text reuse.\\[-2ex]

{\small\noindent
Code, data, and slides: \url{https://webis.de/publications.html\#?q=WWW+2020}}
\end{abstract}

\maketitle

\section{Introduction}

Given a query and a web page, snippet generation is the task of summarizing the web page's content in relation to the query. Today, snippets are basically generated by reusing phrases and sentences from the web page that contain all or at least some of the query's terms. In summarization terminology, this is called {\em extractive summarization}. The alternative {\em abstractive summarization} relaxes the reuse constraint by allowing for abstracting the web page's content by paraphrasing, generalization, or simplification. In this paper, we study the feasibility of query-biased abstractive snippet generation.

\bsfigure[width=\columnwidth]{abstractive-snippets-illustration-final}{Snippet generation paradigms: extractive snippets as done today (top), abstractive snippets (middle) are research in progress, and potential future paradigms may include explained and personalized snippets (bottom).}{Examples of generated snippets.}{Five examples of generated snippets with different properties: abstractive, explanatory, and persoanlized.}

Figure~\ref{abstractive-snippets-illustration-final} contrasts the two paradigms in the context of snippet generation. At the top, a relevant search result for the given query is exemplified, as retrieved by Google at the time of writing; it consists of the result page's title, URL, and an extractive snippet. The snippet is composed of parts of two sentences from the web page, each containing some of the query's terms or synonyms (shown bold). In the middle, two snippets are shown, the first generated with one of our abstractive snippet generators, the second a manually idealized version. None of them appears verbatim on the web page, but they convey very well what the reader will find there. In fact, the abstractive generation approach can be considered more powerful than the extractive generation approach, since snippets generated under the abstractive paradigm give a query-focused page summary instead of presenting loosely or even unconnected text fragments.

As per our previous work \cite{stein:2018b}, another motivation to depart from the extractive snippet generation paradigm can be found in recent EU~legislation, which limits fair use for certain groups of online publishers. Extractive snippets are generated by reusing text segments from original web pages without explicit consent, which arguably might be considered unfair use in cases where reading the snippet instead of visiting the web page suffices to satisfy a need, foreclosing exposure to the page's business model. By contrast, since abstractive snippets are original text, no immediate license fees apply---as long as reuse is avoided. As a goal for future research, we envision abstractive snippets to also be explanatory and personalized, as exemplified at the bottom of Figure~\ref{abstractive-snippets-illustration-final}. Explanations may include reasons for ranking a page high or low, and personalization may hint either at information that the user has not seen elsewhere (i.e., a kind of ``task-progress-based snippet generation'') or at adaptations to the user's experience level on a given subject. 

The generation of abstractive snippets, just like that of abstractive summaries, depends on the availability of large-scale training data. For this purpose the paper introduces two novel approaches to acquire high-quality abstractive snippets at web-scale via distant supervision:
\Ni
anchor contexts, i.e., the text surrounding the anchor text of a hyperlink on a web page, and
\Nii
descriptions from the Directory of Mozilla, DMOZ, the largest open directory project.
Regarding the former, we devised a multi-step mining process that enabled us to extract anchor contexts that are readable, meaningful, and that fulfill the same purpose as abstractive snippets. Regarding the latter, we similarly selected DMOZ descriptions that can be utilized as abstractive snippets. DMOZ qualifies excellently in this regard: The directory contains human-written descriptions for millions of web pages, which succinctly describe usage, purpose, or what users can expect to find on the web pages.

The still-missing piece to render a pair of snippet text (be it an anchor context or a DMOZ description) and web document a suitable training example for abstractive snippet generation is a query. Note that the ideal query must be both relevant for the web document and distinctive for the snippet text. Since we have no control over which web pages are targeted by what snippet text, our approach is to {\em generate matching queries}. Besides the mentioned query generation constraints (relevance and distinctiveness) we consider also web pages that have been judged relevant to the TREC Web, Session, and Task tracks, using their topics as queries. Altogether we were able to construct 3.5~million triples of the type $\langle$query, anchor context, document$\rangle$ and 50~thousand triples of the type $\langle$query, web directory description, document$\rangle$. To better understand the effectiveness of our heuristic mining process, we carried out a qualitative assessment via crowdsourcing.

Our contributions can thus be summarized as follows:
\Ni
We identify the first two sources of ground truth for abstractive snippet generation from which we compile different training corpora (Section~\ref{abstractive-snippet-corpus}).
\Nii
For snippet generation, we utilize pointer-generator networks, which we extend to generate snippets from two directions to ensure that the query words explicitly occur in the generated text (Section~\ref{abstractive-snippet-generation}).
\Niii
We evaluate the training data, as well as variants of our models and competitive baselines via intrinsic and extrinsic evaluation.
The former focuses on the language quality and correctness of the generated text, the latter analyzes summarization effectiveness and snippet usefulness (Section~\ref{evaluation}).

\section{Related Work}

Snippet generation is a variant of the generic summarization problem, where the summaries are concise (2-3 sentences) and biased toward queries. From the start~\cite{luhn:1958}, snippet generation has almost unanimously been based on extractive summarization approaches.

Extractive snippet generation has been subject to research for decades and is the de facto standard for virtually all web search engines~\cite{brin:98}. \citet{tombros:1998} ascertain the importance of query bias in snippets to maximize the accuracy and speed of end users in picking relevant results. Biasing a snippet towards a search query is straightforward: Every sentence in the document is scored based on the distribution of query terms in it, and this score is added to the sentence score obtained by plain extractive summarization methods, denoting the final importance of a sentence as a candidate for the snippet. \citet{white:2002a,white:2002b} give evidence that snippets should be re-generated based on implicit relevance feedback, e.g., when a user returns to a search result page, supporting our vision of personalized snippets. In terms of presentation, \citet{maxwell:17} showed that displaying long, highly informative extractive summaries does not significantly improve the accuracy of the users in selecting relevant results. \citet{kaisser:08} found that snippet length should differ dependent on the type of query, providing evidence that users are able to predict the appropriate snippet length and that their perceptions of search result quality are affected accordingly. \citet{kim:17} further explored the impact of snippet length on mobile devices, concluding that two to three sentences are ideal. Today's deep neural language models are quite capable of generating texts of this length, and, in an extensive user study, we have already shown that manually written abstractive snippets work just as well as extractive ones~\cite{stein:2018k}.

However, despite the recent progress in abstractive summarization spearheaded by \citet{rush:2015}, abstractive snippet generation has received little attention so far, which can be attributed to the lack of a suitable ground truth: \citet{hasselqvist:2017} derived 1.3~million $\langle$query, summary, document$\rangle$-triples from the CNN/Daily Mail corpus~\cite{hermann:2015}, using entities from the summary sentences as queries. Despite the high number of ground truth triples, the summary length of 14~words on average is short compared to the usual snippet length (2-3~sentences), and the text genre is exclusively news. Moreover, despite the fact that the CNN/Daily Mail corpus is the most widely used training dataset for abstractive summarization, its summaries consist mostly of extractive reuse~\cite{stein:2019x}. \citet{nema:2017} created a small corpus consisting of 12,695~triples for a form of query-biased abstractive summarization from Debatepedia. Besides its small size, the corpus comprises only questions with an average length of about 10~words as queries, the average document length is only 66~words, and that of the reference summaries is 11~words. Further afield, \citet{baumel:2018} study query-biased abstractive multi-document summarization based on the DUC~corpora~\cite{dang:2005,dang:2006,dang:2007}, generating summaries of 250~words length. None of these datasets are directly applicable to abstractive snippet generation. Our corpus is not only much larger (3.5~million query-biased triples), it is also not limited to a single genre. Based on our approach to derive queries from document-summary pairs, newly published, true abstractive summarization corpora, such as the Webis-TLDR-17~\cite{stein:2018za,stein:2019x}, can be converted into suitable ground truth as well.

In terms of conditional text generation, the aforementioned works rely on neural networks. \citet{hasselqvist:2017} use a unidirectional query encoder in combination with a bidirectional document encoder to induce query bias. At each decoding time step, the query representation is fed to the decoder as additional input to bias the generated output towards the query. In addition to a query encoder, \citet{nema:2017} use a query attention mechanism, which helps the model focus on parts of the query along with the document. \citet{baumel:2018} also incorporate an attention mechanism into their sequence-to-sequence architecture, projecting sentence-level relevance scores to all words of the sentence, and multiplying them to the attention scores at each decoding time step. Note that all of these approaches aim to induce query bias into their summaries, however, they only do so {\em implicitly} through a dedicated encoder and query attention. A frequent issue with this approach is that, despite these efforts, the generated summaries do not always contain any of the original query terms. To overcome this problem, we {\em explicitly} induce query bias by using query terms as first step of our bidirectional decoder when generating snippets. This way, the query terms are guaranteed to occur in a generated snippet.

\begin{table}[t]%
\small%
\centering%
\setlength{\tabcolsep}{1pt}%
\setuldepth{The}%
\caption{Example of an anchor context as training snippet. The anchor text that linked to the document is highlighted.}%
\label{table-anchor-context-example}%
\begin{tabular}{@{}l@{}}
\toprule
\addlinespace
\parbox{\columnwidth}{{\bfseries Query:} Treasury of Humor} \\[1ex]
\midrule
\addlinespace
\bfseries Snippet: anchor context \\[0.5ex]
\parbox{\columnwidth}{\raggedright Asimov, on the other hand, proposes (in his first jokebook, {\color{blue}\ul{Treasury of Humor}}) that the essence of humour is anticlimax: an abrupt change in point of view, in which trivial matters are suddenly elevated in importance above those that would normally be far more important.} \\
\addlinespace
\midrule
\addlinespace
\bfseries Document \\[0.5ex]
\parbox{\columnwidth}{[\,\textellipsis] Treasury of Humor is unique in that in addition to being a working joke book, it is a treatise on the theory of humor, propounding Asimov's theory that the essence of humor is an abrupt, jarring change in emphasis and/or point of view, moving from the crucial to the trivial, and/or from the sublime to the ridiculous [\,\textellipsis]} \\
\addlinespace
\bottomrule
\end{tabular}%
\end{table}

The evaluation of generated snippets is another key component of the abstractive snippet generation pipeline. \citet{jones:1995} distinguish two fundamental types of evaluation in natural language processing: intrinsic evaluation (the performance of the system) and extrinsic evaluation (the success of achieving a goal). In the context of natural language generation, \citet{gatt:2018} discussed the popular metrics and methods to evaluate generated texts. They suggest that, in order to draw proper conclusions about the effectiveness of a proposed generation approach, it is crucial to report performance across multiple evaluation metrics. This is doubly important, since, as \citet{amidei:2018} observe, there is a general lack of a standardized evaluation framework for generated texts. Thus, employing a variety of approaches maximizes future comparability. In our paper, we extensively evaluate our corpus and the performance of our models in both intrinsic and extrinsic experiments. With respect to the former, we study the quality of the training examples and that of the generated snippets; with respect to the latter, we study if they meet the goals of abstractiveness, absence of reuse, and if they support in selecting relevant results.

\section{Abstractive Snippet Corpus}
\label{abstractive-snippet-corpus}

For the construction of the Webis Abstractive Snippet Corpus~2020,%
\footnote{Corpus: \url{https://webis.de/data.html\#webis-snippet-20}}
we identified the first sources of ground truth for abstractive snippets: anchor contexts and web directories. Our mining pipeline creates the corpus automatically from scratch, given a web archive as input.%
\footnote{Code: \url{https://github.com/webis-de/WWW-20}}
Corpus quality is assessed via crowdsourcing.

\subsection{Anchor Contexts as Abstractive Snippets}

The first surrogate we consider for genuine abstractive snippets are \emph{anchor contexts}. An anchor context is the text surrounding the anchor text of a hyperlink on a web page (see Table~\ref{table-anchor-context-example}). Ideally, it explains what can be found on the linked web page, e.g., by summarizing its contents. The author of an anchor context personally describes the linked web page, enabling readers to decide whether to visit it or not, just like snippets on search results pages. To identify useful anchor contexts that are fluent, meaningful, and close to this ideal, we employ a multi-step mining process. Table~\ref{table-anchor-context-mining} overviews corresponding mining statistics.

\bslabel{Crawling Raw Anchor Contexts}
We mine anchor contexts from the ClueWeb09 and the ClueWeb12 web crawls,%
\footnote{See \url{https://lemurproject.org/clueweb09/}~~and~~\url{https://lemurproject.org/clueweb12/}}
focusing on their 1.2~billion English web pages (500~million from the ClueWeb09 and 700~million from the ClueWeb12). For every hyperlink, we extract its anchor text and 1500~characters before and after as anchor context, trading off comprehensiveness and size of the resulting data. The extracted raw 18~billion and 13~billion anchor contexts, respectively, have been fed into the following nine-step pipeline.

\bslabel{Step 1: Intra-site links}
We assume that anchor contexts of cross-site links are more likely genuine pointers to important additional information compared to intra-site links: The vast majority of the latter are found in menus, footers, buttons, and images, entirely lacking plain text context. We discard all anchor contexts of intra-site links by matching the second-level domain names of the web page containing a given context with that of the linked page. More than 96\%~of the raw anchor contexts are thus removed in this step.

\bslabel{Step 2: Non-existing pages}
We discard anchor contexts that link to pages that are not available in the ClueWeb collections; most of them are meanwhile dead links on the live web. This pertains to~75\% and~82\% of the remaining anchor contexts.

\bslabel{Step 3: Non-English pages}
All anchor contexts whose (linked) page is non-English are discarded. We rely on the language identification done for ClueWeb09 encoded in its document~IDs, whereas the ClueWeb12 is advertised as English-only collection.

\bslabel{Step 4: Spam anchors}
The Waterloo spam ranking provides spam scores for the ClueWeb09 and the ClueWeb12~\cite{cormack:2011}. As suggested, we remove anchor contexts whose (linked) pages' spam rank is~$<$\,70\%. However, we make an exception for anchor contexts whose linked pages have a relevance judgment from one of the TREC Web tracks (2009-2014), Session tracks (2010-2014), or Tasks tracks (2015,~2016).

\begin{table}[tb]%
\small%
\centering%
\setlength{\tabcolsep}{4.5pt}%
\renewcommand{\arraystretch}{0.985}%
\caption{Statistics of the anchor context mining pipeline.}%
\label{table-anchor-context-mining}%
\begin{tabular}{@{}lrrrr@{}}
\toprule
\bfseries Mining pipeline & \multicolumn{2}{@{}c@{}}{\bfseries ClueWeb09}   & \multicolumn{2}{c@{}}{\bfseries ClueWeb12} \\
\cmidrule(l@{\tabcolsep}r@{\tabcolsep}){2-3}\cmidrule(l@{\tabcolsep}){4-5}
& Remaining & \multicolumn{1}{@{}c@{}}{$\Delta$} & Remaining & \multicolumn{1}{c@{}}{$\Delta$} \\
\midrule
Raw anchor contexts      & 17,977,415,779 &          & 12,949,907,331 &          \\
1. Intra-site links      &    440,605,425 &  -97.6\% &    514,337,093 &  -96.0\% \\
2. Non-existing pages    &    111,082,494 &  -74.8\% &     91,007,214 &  -82.3\% \\
3. Non-English pages     &    107,819,314 &   -2.9\% &     91,007,214 &   -0.0\% \\
4. Spam anchors          &     24,767,468 &  -77.0\% &     19,829,007 &  -78.2\% \\
5. Stop anchors          &     17,188,286 &  -30.6\% &     15,837,168 &  -20.1\% \\
6. Improper text         &      9,631,489 &  -44.0\% &      9,248,806 &  -41.6\% \\
7. Duplicated            &      6,292,317 &  -34.7\% &      5,403,893 &  -41.6\% \\
8. Text reuse            &      6,183,783 &   -1.7\% &      5,349,610 &   -1.0\% \\
9. Short web pages       &      5,651,649 &   -8.6\% &      5,114,479 &   -4.4\% \\
\midrule                                                                 
\bfseries Unique pages:  & 2,499,776 & \multicolumn{1}{@{}c@{}}{\ \ --} & 1,557,330 & \multicolumn{1}{@{}c@{}}{\quad--} \\
\bottomrule
\end{tabular}%
\end{table}

\enlargethispage{0.5\baselineskip}
\bslabel{Step 5: Stop anchors}
Anchor contexts whose anchor text is empty, or contains the words ``click'', ``read'', or ``mail'' are removed, since they led our models astray. We also remove multi-link anchor contexts to avoid ambiguous contexts not related to an individual link. As a heuristic, we require a minimum distance of 50~characters between two anchor texts, removing all others.

\bslabel{Step 6: Improper text}
To remove anchor contexts with improper text, we only keep those where
\Ni
the anchor text has at most 10~words (in pilot studies, longer anchor texts were hardly informative or resulted from HTML parsing errors),
\Nii
the anchor text is part of a longer text of at least 50~words (longer texts are a key indicator of meaningful and readable texts),
\Niii
the sentence containing the anchor text has at least 10~words (longer sentences more often resulted in meaningful anchor contexts),
\Niv
the anchor context contains at least one verb as per the Stanford POS tagger~\cite{toutanova:2003}, and
\Nv
the anchor context has a stop word ratio between~10\% and~70\% as per \citeauthor{biber:1999}'s~\cite{biber:1999} study of written English.

\bslabel{Step 7: Duplicated anchor contexts}
To avoid any training bias resulting from duplication, we remove duplicate anchor contexts linking to the same page from different pages. To quickly process all the pairs of anchor contexts for each individual page, we use locality-sensitive hashing~(LSH)~\cite{rao:2016}. We first encoded all anchor contexts as 128-dimensional binary vectors based on word unigrams, bigrams, and trigrams and then removed one of the anchor contexts as ``duplicate'' to another if the cosine similarity of their vectors was larger than~0.9 (this value was determined in pilot studies). Another 34-42\% of the anchor contexts were removed as duplicates.

\bslabel{Step 8: Text reuse}
Since our goal is {\em abstractive} snippet generation, we exclude all anchor contexts that are purely {\em extractive}. This was done by checking if the anchor context was completely copied from their respective linked pages. Partial reuse, i.e., due to reordering of phrases, however, has been retained as a mild form of abstraction.

\bslabel{Step 9: Short web pages}
Finally, we removed anchor contexts whose linked web pages contained less than 100~words to ensure a sufficient basis for summarization. Arguably, snippets need to be generated for shorter pages, too. However, we envision different, specialized snippet generators for different length classes of web pages, which we leave for future work.

Altogether, we obtained 10,766,128 $\langle$anchor context, web docu\-ment$\rangle$ tuples from the two ClueWeb collections referring to 4,057,106 unique pages. The average length of an extracted anchor contexts is 190~words (longest:~728; shortest:~50). The average linked page is 841~words long (longest:~14,339; shortest:~100) and it has 2.65~anchor contexts, while about two-thirds (2,675,980~pages) have only one, and the most often linked page has 12,925 anchor contexts.

\subsection{DMOZ Descriptions as Abstractive Snippets}

\begin{table}[tb]%
\small%
\centering%
\setuldepth{The}%
\caption{Example of a DMOZ description as training snippet.}%
\label{table-dmoz-example}%
\begin{tabular}{@{}l@{}}
\toprule
\addlinespace
\parbox{\columnwidth}{{\bfseries Query}: Customer Respect Index} \\[1ex]
\midrule
\addlinespace
\bfseries Snippet: DMOZ description \\[0.5ex]
\parbox{\columnwidth}{\raggedright {\color{blue}\ul{The Customer Respect Group}}: An international research and consulting firm, publishes the Online Customer Respect Index (CRI) and provides industry and company-specific research and analysis to help companies increase sales and customer retention by improving how they treat their customers online.} \\
\addlinespace
\midrule
\addlinespace
\bfseries Document \\[0.5ex]
\parbox{\columnwidth}{[\,\textellipsis] The Customer Respect Group has been a trusted source of online benchmark data and strategic insight since 2003. While much of our work is in financial services, we have worked across a variety of industries including telecommunications, education, government, and retail. [\,\textellipsis]} \\
\addlinespace
\bottomrule
\end{tabular}%
\end{table}

The second surrogate we consider for genuine abstractive snippets are {\em web directory descriptions}. Web directories used to be a key resource for search and retrieval in the early days of the web. However, the rise of web search engines, and ultimately the success of Google, heralded their slow demise between 2000 and~2010.

Despite the many web directories that have been in operation, few survived to this day. The shutdown of the directories that have been operated by basically all major companies in the search market (even Google) have rendered them permanently unavailable. The one we could still obtain for our purposes is the well-known ``Directory of Mozilla'' (DMOZ), one of the largest open source web directories in its time. Each of its more than three million web pages came with a short, usually one or two sentences long, human-written description. Table~\ref{table-dmoz-example} shows an example. Because the DMOZ descriptions are well-written and because they explain what users would find on the linked website (individual pages were found less often in web directories), they can be considered high-quality summaries with a high level of abstraction. Unfortunately, not only the original DMOZ website, but also most crawls of DMOZ that have been compiled at different points in time, are unavailable today. We found one published at Mendeley,%
\footnote{\url{https://data.mendeley.com/datasets/9mpgz8z257/1}}
and another through the Internet Archive's Wayback Machine.%
\footnote{\fontsize{6.7pt}{7pt}\selectfont\url{https://web.archive.org/web/20160306230718/http://rdf.dmoz.org/rdf/content.rdf.u8.gz}}

We were able to retrieve a 3,200,765 $\langle$web page description, URL$\rangle$ tuples from these crawls, but given their age, most of the URLs are not accessible, anymore. We crawled the ones that still exist today, and extracted their contents, obtaining 574,720 $\langle$DMOZ description, document$\rangle$ tuples. Given the a priori high quality of the anchor contexts, we did not run them through the entire mining pipeline we used for anchor contexts, but only applied Step~3 (by using an automatic language identifier), Step~6, and Step~9. The average length of the extracted DMOZ descriptions is 59~words (longest:~159; shortest:~50) with only one description per web document, where the average document is 241~words long (longest:~42,101, shortest:~100). We believe that a more complete recovery of DMOZ and its linked web pages is possible through the Wayback Machine, which we leave for future work.

\subsection{Query Generation}

The final step of constructing our corpus was query generation for the two sets of $\langle$anchor context, web document$\rangle$ tuples and $\langle$web directory description, web document$\rangle$ tuples. While theses sets of tuples can already be used as ground truth for generic abstractive summarization (which we do as part of our experiments), they are still unsuited to train a {\em query-biased} abstractive snippet generation model for lack of a query. To be suitable training examples, we require for every tuple a query for which
\Ni
the document is (at least marginally) relevant, and
\Nii
the abstractive snippet surrogate is (at least marginally) semantically related.

One might consider the anchor text (i.e., the actually hyperlinked text) to be a suitable candidate for a query that sufficiently fulfills these constraints. A cursory analysis, however, suggested that the texts the authors of the anchor contexts chose to include in their links are not necessarily well-suited for our purposes, even when excluding stop anchors containing words like `click' and `here' as per Step~5 of our anchor context mining pipeline. To avoid this potential for bias, we instead resorted to keyphrase extraction from the entire anchor context to generate queries. Regarding the web directory descriptions, this was the only alternative, anyway.

First, for each tuple, we parse the abstractive snippet surrogate and its associated document using the Stanford POS tagger~\cite{toutanova:2003} and extract all noun phrases with a maximum length of six words. Here, we apply the more limited definition of strict noun phrases by \citet{hagen:2012}, where a strict noun phrase has only adjectives, nouns, and articles. This maximizes tagging reliability and limits the types of queries we consider. Second, to ensure a strong semantic connection between anchor context and document, we use only those noun phrases as queries that appear in both the abstractive snippet surrogate as well as the document. We generate at most three queries per tuple with an average length of 2.43~words (longest:~6; shortest:~1).

The additional constraint that the extracted query has to occur in both the abstractive snippet surrogate as well as the linked document limits the usable tuples. In the end, we obtained a total of 3,589,701 $\langle$query, anchor context, document$\rangle$ triples and 55,461 $\langle$query, web directory description, document$\rangle$ triples corresponding to~33.4\% and~9.7\% of the respective original sets of tuples.

The constraints we impose on the shape of queries and their relation to the abstractive snippet surrogates and the document may seem extreme. However, we argue that a tight control of the texts, and their relation to the query is crucial due to the noisy web data. We leave the study of relaxing these constraints and studying other types of queries (e.g., questions) for future work.

\subsection{Web Content Extraction}

Web pages are a notoriously noisy source of text: A naive approach to content extraction, e.g., by simply removing all HTML markup, is frequently found to be insufficient. Instead, we adopt the content extraction proposed by \citet{kiesel:2017}. Here, the web page is first ``rendered'' into a plain text format, so that blocks of text can be discerned. Then, noisy text fragments, such as menus, image captions, etc., are heuristically removed. This includes paragraphs with less than 400~characters, sentences with less than~50\% letter-only tokens, and sentences without an English function word.

\subsection{Quality Assessment}

We employ crowdsourcing to evaluate
\Ni
the quality of the anchor contexts,
\Nii
the quality of the generated queries, and,
\Niii
the quality of the anchor contexts when used directly as query-biased snippets.
We do not evaluate the quality of the DMOZ descriptions, given their high a priori quality.

We selected 200~$\langle$query, anchor context, document$\rangle$ triples to be assessed. Among them, 100~triples (Group~A) have documents which received relevance scores of~2 within the TREC Web, Session, or Task tracks, which means that these documents have been considered highly relevant to a given TREC topic. In this group, we use the queries of the TREC topics to examine our query generation in relation to the ones supplied for the tracks. Another 100~triples (Group~B) have received no relevance scores from TREC. For all the three crowdsourcing studies, each task was done by five workers. We calculated the average score for an anchor context based on the following annotation scheme: \emph{very bad} gets score~-2; \emph{poor} gets score~-1; \emph{okay} gets score~1; \emph{very good} gets score~2.

First, in the study of anchor context quality, workers were shown individual anchor contexts to validate that the anchor contexts remaining after our pre-processing steps have a high language quality. On average, the quality score is~1.06 (1.09~in Group~A and~1.02 in~B). The scores show that the quality of the anchor contexts can be expected to be \emph{okay}. Besides, there is no difference between the two groups ($p$-value of $t$-test is~0.46). Because the two groups are selected unrelated to anchor context quality, no such differences can be expected when judging them.

Next, the annotators judged if the generated queries are important with respect to their respective anchor contexts to validate our query generation approach. The average query quality score is~0.28 (0.15~in Group~A and~0.41 in~B), showing the overall query quality is just above average. The difference between Group~A and~B clearly shows that the queries from Group~A (obtained from TREC topics) do not fit the anchor contexts, whereas they do fit to the documents.

Lastly, we study if the anchor contexts can be used directly as query-biased snippets by showing the entire triple to the workers. Here, the average score is~-0.08, underlining that the anchor contexts may allow for distantly supervised training, but not close supervision. The average score of Group~A is~-0.68, but that of Group~B is~0.52, which is further evidence that the generated queries better match the anchor contexts than those of the TREC topics.

Altogether, the three crowdsourcing studies have given us confidence that the anchor contexts we mined are reasonably well-suited to serve as summaries of their linked web documents, and that the queries generated for them serve as a reasonable point of connection between them. By extension this also applies to the DMOZ descriptions, since high writing quality can be presumed here.

\section{Abstractive Snippet Generation}
\label{abstractive-snippet-generation}

Our main approach to generate abstractive snippets used our corpus of $\langle$query, snippet, document$\rangle$ triples for training and is an adaptation of pointer-generator networks for query-biased snippet generation. In addition to comparing two variants of our model, we also consider four baselines, including one state-of-the-art abstractive summarization model, and one extractive summarizer with a paraphraser attached.

\subsection{Pointer-Generator Network with Bidirectional Generation}

Pointer-generator networks with copy mechanism~\cite{see:2017} are one of the most successful neural models used for sequence-to-sequence tasks, such as summarization and machine translation. They tackle two key problems of basic sequence-to-sequence models: reproducing facts and handling out-of-vocabu\-lary words. At each time step, the decoder of this model computes a generation probability that decides whether to generate the next word or whether to copy a word from the to-be-summarized text. This provides a balance between extractive and abstractive aspects during the generation process, since, even in the abstractive summarization scenario, reusing some phrases/words is still an essential feature to preserve information from the source text that cannot be abstracted. For instance, named entities should not be exchanged. We note that this aligns with our aim of generating factually correct abstractive snippets with limited text reuse. For our experiments, we use the PyTorch implementation provided by OpenNMT~\cite{klein:2017}.

\bsfigure[scale=0.55]{abstractive-snippet-generator}{Schematic of our abstractive snippet generator.}{Diagram of our snippet generation model.}{Diagram of our snippet generation model that depicts a sequence to sequence setup with two decoders, generating to the beginning and the end of a snippet starting with the query terms.}

Although pointer-generator networks have been shown to generate abstractive summaries reasonably well, they are not designed for generating query-biased abstractive snippets, nor to {\em explicitly} insert designated terms into a generated summary; the expectation for snippets on search engines is to at least include some of the query's terms or synonyms thereof. We therefore devise a pointer-generator network that guarantees by construction that the query occurs in an abstractive snippet.

Our \emph{bidirectional generation model} generates an abstractive snippet using two decoders as depicted in Figure~\ref{abstractive-snippet-generator}. We set up the query words to be the first words in their output and then the model learns to complete the snippet in two directions: starting from the query to the end of the snippet, and starting from the query to its beginning. Formally, given a query~$q$, a document~$d$, and the target snippet $s = \langle s_\mathrm{prev}, q, s_\mathrm{next} \rangle$, where $s_\mathrm{prev}$ ($s_\mathrm{next}$) is the snippet before (after) the query~$q$, our model trains an encoder~$E$ to encode the concatenation of query~$q$ and document~$d$ to a context vector~$z \sim E(q,d)$. Furthermore, two decoders~$D_\mathrm{prev}$ and~$D_\mathrm{next}$ generate the snippet from vector~$z$: one from the query's terms to the beginning of the snippet~$\hat{s}_\mathrm{prev} \sim D_\mathrm{prev}(z,q)$; the other from the query's terms to the end of the snippet~$\hat{s}_\mathrm{next} \sim D_\mathrm{next}(z,q)$. The final generated snippet is the concatenation of~$\hat{s}_\mathrm{prev}$,~$q$, and~$\hat{s}_\mathrm{next}$.

\subsection{Input Preparation}

Our model is trained with the $\langle$query, snippet, document$\rangle$ triples from our corpus. Given the fact that the query's terms can occur anywhere in the document (even multiple times), and given the wide range of document lengths versus the limited input size of our model, we resort to an initial extractive summarization step as a means of length normalization while ensuring that the document's most query-related pieces of text are encoded in first place. This, however, may negatively affect fluency across sentences. We use the query-biased extractive snippet generator by \citet{liu:2018}, which computes query-dependent TF-IDF weights to rank a document's sentences, and then selects the top~10 as input, truncating at 500~words.

\subsection{Model Variants and Baselines}

We train two variants of our model, one using anchor context triples ({\small\tt AnCont.-QB}), and another one using DMOZ triples ({\small\tt DMOZ-QB}). These models generate query-biased snippets as described above.

Furthermore, we consider four baselines. The first two baselines employ our model as well, but without enforcing query bias: we train one model on the anchor context triples ({\small\tt AnCont.}) and another on the DMOZ triples ({\small\tt DMOZ}) without using the query as predefined output. Thus, we can simplify our model to using just a single decoder. This way, there is a chance for the models to learn to generate query-biased snippets by themselves, however, for each generated snippet, there is also a chance that they will end up not being query-biased.

Another two baselines allow for a comparison with the state of the art in abstractive summarization and paraphrasing. The former is a single layer Bi-LSTM model trained on the CNN/Daily Mail corpus for abstractive summarization ({\small\tt CNN-DM}).%
\footnote{\url{http://opennmt.net/Models-py/}}
This model implements a standard summarizer and will therefore not generate query-biased summaries. The paraphrasing baseline ({\small\tt Paraphr.}) is a way of combining conventional extractive snippet generators with paraphrasing technology to achieve the same goal of producing snippets that are abstractive and that do not reuse text. This baseline operationalizes the manual creation of paraphrased snippets as in our previous user study~\cite{stein:2018k}, albeit with a lower text quality as paraphrasing models are still not perfectly mature. We first create a query-biased extractive summary of the web page consisting of three sentences~\cite{liu:2018}, and then apply the pre-trained paraphrasing model of \citet{wieting:2017}.%
\footnote{\url{https://github.com/vsuthichai/paraphraser}}

\section{Evaluation}
\label{evaluation}

We carried out both an intrinsic and an extrinsic evaluation of the generated snippets for all model variants and the baselines. While the intrinsic evaluation examines the language quality of the generated snippets, the extrinsic evaluation assesses their usefulness in the context of being used on a search engine. For both intrinsic and extrinsic evaluation, we carry out crowdsourcing experiments employing master workers from Amazon Mechanical Turk with a minimum approval rate of~95\% and at least~5000 accepted HITs (Human Intelligence Task).%
\footnote{We paid an average hourly rate of~\$5.60 and a total amount of~\$622.50.}  

Table~\ref{table-split} shows our corpus split into training, validation, and test sets. Note that the two baseline models {\small\tt CNN-DM} and {\small\tt Paraphraser} can be trained on the entirety of snippet-document pairs extracted from the ClueWeb collections and DMOZ, regardless whether queries can be generated for them. From both the DMOZ descriptions and the anchor contexts, we randomly selected 1000 documents for validation. This resulted in 1000~DMOZ triples because of the one-to-one correspondence between DMOZ descriptions and linked documents, but 3842~anchor context triples. Likewise, for testing, we selected another 1000~documents each, resulting in 3894~anchor context triples.

\begin{table}[tb]%
\centering%
\small%
\renewcommand{\tabcolsep}{4pt}%
\caption{The number of instances for training, validation, and testing.}%
\label{table-split}%
\begin{tabular}{@{}lrcc@{}}
\toprule
\bfseries Model & \bfseries Training & \bfseries Validation & \bfseries Test\\
\midrule
DMOZ-QB           &     54,461 & 1,000 & 3,894\\
AnchorContext-QB  &  3,581,965 & 3,842 & 3,894\\
\midrule
DMOZ              &    573,720 & 1,000 & 3,894\\
AnchorContext     & 10,758,392 & 3,842 & 3,894\\
\bottomrule
\end{tabular}%
\end{table}

\subsection{Intrinsic Evaluation}
\label{intrinsic-evaluation}

We computed ROUGE F-scores~\cite{lin:2004} for all the models as seen in Table~\ref{table-evaluation-results}a. The ROUGE scores are computed by comparing the n-gram overlap between the snippets generated by each model and the reference snippets in the test set. Among the two baselines {\small\tt CNN-DM} and {\small\tt Paraphraser}, {\small\tt CNN-DM} achieves higher ROUGE scores across all granularities. Because {\small\tt CNN-DM} was trained to summarize articles while {\small\tt Paraphraser} only paraphrases the first three sentences from an extractive summary, it is unsurprising that snippets generated by {\small\tt CNN-DM} fit the gold standard better. Also, we observe that inducing query bias by reshaping the corpus leads to better ROUGE scores for both DMOZ descriptions and anchor contexts. The {\small\tt AnchorContext-QB} model has the best performance among all variations. It shows that the model can successfully generate snippets close to the gold standard with help from the bidirectional architecture. However, notwithstanding the smaller training corpus of 54,461 instances, we also see the effectiveness of this approach in the {\small\tt DMOZ-QB} model.

Next, we assessed fluency, factuality, and text reuse of the snippets with respect to the to-be-summarized document. We selected 100~documents from the test set and corresponding snippets generated by each model. The size of the test set is reduced as we also employed manual evaluation alongside automatic evaluation. Also, we ask the human judges to annotate only high-quality examples that were chosen using their perplexity scores as mentioned below to showcase the potential of abstractive snippet generation.

\bslabel{Fluency}
Calculating fluency automatically can be done using a language model where the perplexity score implies how fluent (probable) a text is, with lower perplexity for more fluent texts. We used a publicly available BERT model~\cite{devlin:2019} to compute perplexity scores for the snippets.%
\footnote{We used \emph{BERT-Large, Uncased} for our experiments.}
These perplexity scores were also used to select the test set for manual evaluation. 

The average perplexity scores of the our models and the four baselines are shown in the first column of Table~\ref{table-evaluation-results}b. {\small\tt CNN-DM} and {\small\tt AnchorContext} generate fluent texts with low perplexities. While {\small\tt CNN-DM} is trained on well-written news articles with extractive summaries~\cite{grusky:2018}, the high performance (low perplexity) of the {\small\tt Anchor\-Context} model can be attributed to the relatively large corpus of 10~million training examples. However, in case of {\small\tt AnchorContext-QB}, just the addition of query bias in the snippet generation process introduces breaks in the text flow, thereby introducing a small increase in the perplexities. {\small\tt DMOZ}'s perplexity is similar to {\small\tt AnchorCon\-text} or {\small\tt AnchorContext-QB}, showing that the DMOZ descriptions' language fluency is pretty high. In case of {\small\tt DMOZ-QB}, the poor performance can be attributed to a strong repetition of tokens (sometimes more than 10~times) in the generated snippets. Besides, we also see a pretty high perplexity for {\small\tt Paraphraser}, which implies that simply rephrasing words without considering the whole context largely reduces the fluency of texts.

\begin{table*}[tb]%
\centering%
\small%
\renewcommand{\tabcolsep}{2pt}%
\caption{Evaluation results:
(a)~Snippet overlap with the ground truth.
(b)~Snippet quality: fluency is measured as perplexity (lower is better), factuality as strict noun phrase overlap (higher is better), and reuse as ROUGE-L precision (lower is better).
(c)~Snippet fluency,
(d)~summarization effectiveness, and
(e)~query bias as judged by crowd workers, where each study is based on $n=100$ snippets, three votes each, and agreement is measured as 2/3~majorities and full agreement.
(f)~Usefulness of snippets to crowd workers in selecting relevant documents, compared to \citeauthor{stein:2018k}'s~\cite{stein:2018k} study of manually paraphrased snippets.}%
\label{table-evaluation-results}%
\begin{tabular}[t]{@{}l@{}}
\\
\toprule
\textbf{Model} \\
\addlinespace
\addlinespace
\\
\\
\midrule
CNN-DM         \\
Paraphraser    \\
DMOZ           \\
DMOZ-QB        \\
AchrCtxt.      \\
AchrCtxt.-QB   \\
\midrule
\citet{stein:2018k} \\
\bottomrule
\end{tabular}
\hfill%
\begin{tabular}[t]{@{}ccc@{}}
\multicolumn{1}{@{}l@{}}{(a)} \\
\toprule
\multicolumn{3}{@{}c@{}}{\bfseries ROUGE} \\
\midrule
 1 & 2 & L  \\
\\[1.25ex]
\midrule
    20.5 &     4.2  &     15.4  \\
    14.3 &     1.1  &     10.5  \\
    15.6 &     1.7  &     11.4  \\
    20.8 &     2.8  &     15.6  \\
    23.0 &     5.0  &     18.0  \\
\bf 25.7 & \bf 5.2  & \bf 20.1  \\
\bottomrule
\end{tabular}%
\hfill%
\begin{tabular}[t]{@{}rrr@{}}
\multicolumn{1}{@{}l@{}}{(b)} \\
\toprule
\multicolumn{3}{@{}c@{}}{{\bfseries Quality} (autom.)}  \\
\midrule
Fluency & Fact. & Reuse  \\
\\[1.25ex]
\midrule
       2.42 & \bf 76.10 &     83.17 \\
     422.20 &     45.21 &     68.61 \\
       2.59 &      2.02 &     33.75 \\
    1215.18 &      6.89 & \bf 29.55 \\
\bf    2.04 &      6.19 &     30.37 \\
       2.31 &     17.89 &     45.36 \\
\bottomrule
\end{tabular}%
\hfill%
\begin{tabular}[t]{@{}ccrrrrr@{}}
\multicolumn{1}{@{}l@{}}{(c)} \\
\toprule
\multicolumn{7}{@{}c@{}}{{\bfseries Fluency} (crowd)} \\
\midrule
\multicolumn{2}{@{}c}{Agrmnt.} & \multicolumn{4}{c@{}}{Score} & Avg. \\
\cmidrule(r@{\tabcolsep}){1-2}\cmidrule(l@{\tabcolsep}r@{\tabcolsep}){3-6}
Maj. & Full & -2 & -1 & 1 & 2 \\
\midrule
  86\% & 28\% &      0 &      1 &     24 &     75 & \bf 1.73 \\
  78\% & 22\% &      7 &     17 &     40 &     36 &     0.81 \\
  81\% & 17\% &      2 &      4 &     43 &     51 &     1.37  \\
  97\% & 51\% & \bf 51 & \bf 41 &      7 &      1 &    -1.34 \\
  90\% & 32\% &      0 &      3 &     20 & \bf 77 &     1.71 \\
  73\% & 10\% &      3 &     35 & \bf 44 &     18 &     0.39 \\
\bottomrule
\end{tabular}%
\hfill%
\begin{tabular}[t]{@{}ccrrrrr@{}}
\multicolumn{1}{@{}l@{}}{(d)} \\
\toprule
\multicolumn{7}{@{}c@{}}{{\bfseries Effectiveness} (crowd)} \\
\midrule
\multicolumn{2}{@{}c}{Agrmnt.} & \multicolumn{4}{c@{}}{Score} & Avg. \\
\cmidrule(r@{\tabcolsep}){1-2}\cmidrule(l@{\tabcolsep}r@{\tabcolsep}){3-6}
Maj. & Full & -2 & -1 & 1 & 2 \\
\midrule
  90\% & 20\% &      4 &     15 &     28 & \bf 53 & \bf 1.11 \\
  81\% & 14\% &      4 &     22 &     37 &     37 & 0.81 \\
  70\% & 17\% &     14 & \bf 27 &     37 &     22 & 0.26 \\
  91\% & 40\% & \bf 38 &     16 &      6 &      2 & -0.82 \\
  79\% & 19\% &     11 &     14 & \bf 38 &     37 & 0.76 \\
  75\% & 16\% &      4 &     26 &     32 &     38 & 0.74 \\
\bottomrule
\end{tabular}%
\hfill%
\begin{tabular}[t]{@{}cc@{\hspace{1em}}cc@{}}
\multicolumn{1}{@{}l@{}}{(e)} \\
\toprule
\multicolumn{4}{c@{}}{{\bfseries Qry. Bias} (crwd.)} \\
\midrule
\multicolumn{2}{c@{\hspace{1em}}}{Maj.} & \multicolumn{2}{c@{}}{Full} \\
\cmidrule(r@{1em}){1-2}\cmidrule{3-4}
   Yes &     No &    Yes &     No \\
\midrule
    79 &     21 &     41 &     21 \\
    73 &     27 &     23 &     27 \\
    60 & \bf 40 &     10 & \bf 40 \\
    69 &     31 &     12 &     31 \\
    82 &     18 &     34 &     18 \\
\bf 87 &     13 & \bf 43 &     13 \\
\bottomrule
\end{tabular}%
\hfill%
\begin{tabular}[t]{@{}c@{}}
\multicolumn{1}{@{}l@{}}{(f)} \\
\toprule
\bfseries Use- \\[0.625ex]
\bfseries flns. \\[0.625ex]
\midrule
F \\
\midrule
       61.85 \\
       60.49 \\
       46.03 \\
       59.82 \\
       34.86 \\
 \bf   66.18 \\
\midrule
 \bf 67.64 \\ 
\bottomrule
\end{tabular}%
\end{table*}

In addition to automatically computing perplexity scores, we performed a manual evaluation where we asked workers to score the fluency on a 4-point Likert scale from \emph{very bad} via \emph{bad} and \emph{good} to \emph{very good} corresponding to scores from~-2 to~2. The workers were only presented with the generated snippets in the HIT interface. Table~\ref{table-evaluation-results}c shows the results of this qualitative evaluation. We achieved a high inter-annotator agreement with the lowest majority agreement of 73\%  and the highest one of 97\%. Among all models, {\small\tt CNN-DM} achieves the highest average fluency score, with {\small\tt AnchorContext} performing competitively. {\small\tt DMOZ} also achieves a rather high average score in fluency. Given the comparably small number of DMOZ descriptions available for training than that of anchor contexts, such a reduction of fluency can be expected from the generated snippets. Besides, the scores of query-biased model ({\small\tt DMOZ-QB} and {\small\tt AnchorContext-QB}) are significantly lower than the scores of query-unbiased models ({\small\tt DMOZ} and {\small\tt AnchorContext}). This shows that when our architecture generated snippets with the requirement to explicitly put the query words in the snippets, the model compromised the language fluency in order to meet this requirement.

\bslabel{Factuality}
\citet{cao:2018} showed that neural text generation models can create fluent texts despite conveying \emph{wrong} facts. One reason is that factual units, such as names, locations, or dates, are infrequent in corpora, which leads to their weak representation in the final embedding space. The recently proposed copy mechanism~\cite{gu:2016}, and pointer generator networks~\cite{see:2017} mitigate this problem to some extent. For example, in summarization, models with the copy mechanism learn to reuse some words/phrases from the documents to be summarized and simply copy them to the generated summary.

However, ROUGE cannot be used to evaluate factuality as it does not specifically count factual units while computing the n-gram overlap. Thus, we formulate the factuality evaluation as calculating the ratio of strict noun phrases preserved by the generated snippet for a given document: $|S \cap \hat{S}|/|\hat{S}|$, where $S$ is the set of strict noun phrases in a document, and $\hat{S}$ is the set of strict noun phrases in its generated snippet. Recall that a strict noun phrase is defined as a noun phrase with a head noun and an optional adjective, which have also been exclusively considered for query generation. This ratio approximates the amount of factual units from the document that are preserved by the generated snippet.

The factuality scores can be found in the second row of Table~\ref{table-evaluation-results}b. We see that {\small\tt CNN-DM} and {\small\tt Paraphraser} have a much higher ratio of strict noun phrases than the other models. Manual inspection of the generated snippets reveals excessive copying of text from the document to the snippet by both models. This preserves a large number of factual units, albeit impacting the desired property of abstractiveness in the snippets. Also, this increases the amount of text reuse. As explained in Section~\ref{abstractive-snippet-corpus}, the anchor contexts are relatively abstractive, which makes reproducing facts rather difficult for models trained on our corpus. However, generating query-biased snippets ({\small\tt DMOZ-QB} and {\small\tt AnchorContext-QB}) leads to some improvement of the factuality scores. We attribute this to concatenating the query term and the web page during the training which improves the strict noun phrase overlap. 

\bslabel{Text reuse}
In addition to being fluent and factually correct, an ideal abstractive snippet also avoids text reuse from the document. We enforced this property during corpus construction by filtering out anchor contexts that largely reuse text from the web document. To evaluate the impact of this step, we calculate the amount of text reuse as ROUGE-L precision between the generated snippet (candidate) and the document (reference). A lower precision implies lower text reuse by the generated snippet. The third row of Table~\ref{table-evaluation-results}b shows the results of the automatic evaluation of text reuse. The baselines have very high text reuse, especially {\small\tt CNN-DM}, so that most of the generated snippets are sentences copied from the documents. As the training corpus for {\small\tt CNN-DM} does not contain many abstractive summaries as references, this is not unexpected as the model's copy mechanism gains significantly higher importance during training. Our four model variations exhibit much lower text reuse. Except for {\small\tt AnchorContext-QB}, the other three have similar text reuse rates---about one-third. This shows that our models have learned to balance reuse with abstractiveness.

\subsection{Extrinsic Evaluation}

Our extrinsic evaluation assesses how well the generated snippets can be used in practice. We designed two crowdsourcing tasks to evaluate summarization effectiveness and snippet usefulness.

\bslabel{Summarization Effectiveness}
A usable snippet should ideally describe the web document and help users make an informed decision whether or not to click on a search result returned for the search query. In one HIT, we showed the query, a snippet generated by one the models, and the summarized web page. Workers were asked to score how helpful the given snippet was at describing the document on a four-point Likert scale defined as follows:
\emph{Very poor}~(-2): The snippet does not describe the web page and is useless.
\emph{Poor}~(-1): The snippet has some information from the web page but doesn't help decide to visit the web page.
\emph{Acceptable}~(1): The snippet has key information from the web page and helps decide to visit the web page.
\emph{Good}~(2): The snippet describes the web page really well and helps decide to visit the web page.
The results of this task can be found in Table~\ref{table-evaluation-results}d. We achieved a high inter-annotator agreement with the lowest majority agreement of~75\% and the highest one of~91\%. The table shows that {\small\tt CNN-DM} best summarizes the documents, while {\small\tt Paraphraser} and the anchor context models {\small\tt AnchorContext} and {\small\tt AnchorContext-QB} perform comparably well. We see that {\small\tt DMOZ-QB} has a much lower average score than the others. The low scores are reflected also by the low language fluency of the text and the low factuality as shown in the previous evaluations.

Additionally, we asked workers to judge if the snippet is query-biased when describing the document. This helps us further assess if our query-biased models do in fact generate snippets that consider the search query in their description of the web document. Table~\ref{table-evaluation-results}e shows the full agreements among workers for this specific question. We observe that {\small\tt AnchorContext-QB} and {\small\tt CNN-DM} are the top two models where the snippets are query-biased (43 and~41). However, {\small\tt CNN-DM} also has a higher number of \emph{no} votes. Also, compared to query-unbiased models ({\small\tt DMOZ} and {\small\tt AnchorContext}), the query-biased models ({\small\tt DMOZ-QB} and {\small\tt AnchorContext-QB}) can generate snippets that are more query-focused. It follows that our process of shaping the training examples to be query-biased is successful and our models can learn to generate such snippets. Given the fact that the DMOZ descriptions are often too short to reliably shape them into query-biased training instances, the {\small\tt DMOZ} model fails to generate a high number of query-biased snippets.

\bsfigure[width=\columnwidth]{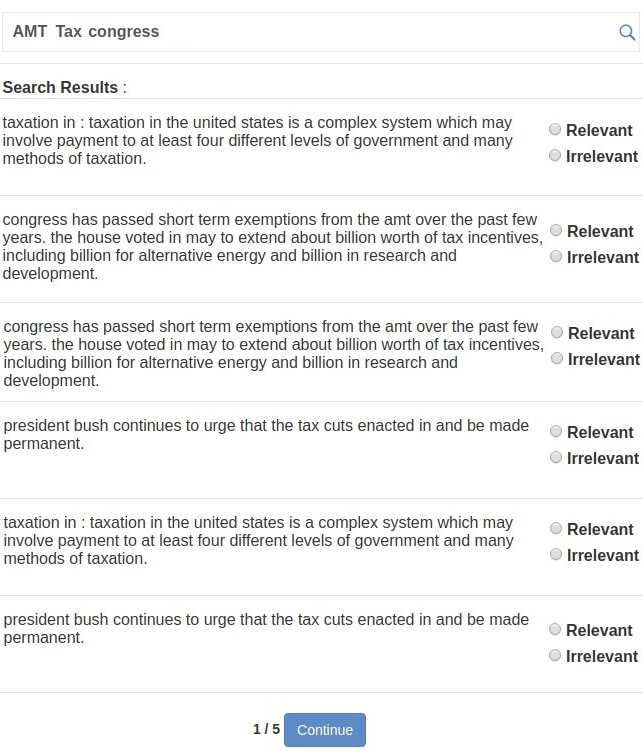}{The HIT interface of the usefulness experiment.}{Worker interface for the snippet usefulness experiment.}{The interface used by workers in deciding snippet usefulness, showing a number of snippets next to radio boxes where the expected relevancy of a document on a search results page is to be predicted.}

\bslabel{Snippet Usefulness}
With this crowdsourcing task we evaluate whether the users of a search engine are able to identify relevant results for a given search query based on the generated snippet of each document. We follow our previously applied experimental setup~\cite{stein:2018k}. First, we selected 50~topics as queries from the aforementioned TREC Web tracks, Session tracks, and Tasks tracks, ensuring that the queries had at least three relevant and three irrelevant documents judged in the datasets provided by the tracks. We evaluated each model independently by showing the search query and snippets of six documents (whose relevance judgments are known) generated by this model. The interface of this annotation task, as presented to the workers, can be seen in Figure~\ref{hit-interface3}, where workers judged each snippet to be relevant or irrelevant. This task emulates the use of abstractive snippets in a practical setting.

Table~\ref{table-evaluation-results}f shows the results of this experiment. For comparison, we show the results of \citet{stein:2018k}, where the extractive snippets as employed by Google achieved the highest F-score of~67.64 among their approaches. The baselines {\small\tt CNN-DM} and {\small\tt Paraphraser} perform similar to each other but are still worse than \citeauthor{stein:2018k}'s extractive snippets. Both DMOZ-based models performed worse than the baselines, while {\small\tt DMOZ-QB} performs a lot better than {\small\tt DMOZ}; its snippets have more query bias. The {\small\tt AnchorContext-QB} model performs competitively to \citeauthor{stein:2018k}'s extractive snippets (66.18~versus~67.64), while comprising significantly less text reuse (see last row of Table~\ref{table-evaluation-results}b). This result is very promising, implying sufficiently abstractive snippets that are useful to identify relevant documents, like the extractive snippets of commercial search engines.

\section{Discussion}

To contextualize our evaluation, in this section, we discuss and analyze an individual example and the shortcomings of the different models. This is followed by our ranking of models.

\subsection{Examples of Generated Snippets}

Table~\ref{table-examples} shows our example query ``cycling tours'' and an excerpt of a relevant document from the ClueWeb09 right below. Each of the models under evaluation has generated a snippet for this document, listed below the excerpt.

The {\small\tt CNN-DM} model copies two sentences from the document (the second one appearing later). This is unsurprising, since the model's training corpus exhibits a strong extractive bias. By accident, one of the sentences contains a query term (``cycling''), rendering the snippet partially query-biased. A common problem with neural text generation is exemplified, namely the repetition of words and phrases (``Vermont''). For training and generation, numbers have been replaced with {\small\tt <num>}. The {\small\tt Paraphraser} model paraphrases the first sentence, accidentally removing the snippet's query bias while introducing terminology related to bicycling (``transmission''), but not related to the document's topic. It generates erroneous statements and has problems with the usage of definite and indefinite determiners. Still, some text remains untouched, so that a little less than half the snippet is reused text. These observations are in line with our evaluation results, where {\small\tt Paraphraser} is found to exhibit lower fluency scores due to less reuse (Tables~\ref{table-evaluation-results}b and~\ref{table-evaluation-results}c) and is less query-biased (Table~\ref{table-evaluation-results}e) than {\small\tt CNN-DM}. Our basic factuality measures do not yet include fact checking, thus overlooking false statements.

Regarding our {\small\tt DMOZ} model, if not for the repetition and a factual error at the end, it would have generated a to-the-point, query-biased snippet. Its style resembles that of DMOZ summaries, quite befitting a snippet (e.g., by omitting the subject in the second sentence). Still, this model achieves the lowest performance in summarization effectiveness (Table~\ref{table-evaluation-results}d) and query bias (Table~\ref{table-evaluation-results}e), although its measured fluency is much higher compared to that of {\small\tt DMOZ-QB}. The latter also generates a rather to-the-point snippet, albeit also with repetitions and improper termination ({\small\tt <eos>}). Starting with the query ``cycling tours'' and generating in two directions, the backward generation to the beginning of the snippet apparently failed, introducing a false company name (``Spring tours cycling tours''). Neither of the DMOZ models reused any phrase longer than two words from the original document.

The two anchor context models provide for the most fluent examples that are not reusing any text verbatim. The {\small\tt AnchorContext} model does not repeat itself and accidentally includes at least one of the query's terms. It introduces a company not referred to in the document, and it introduces bike tours at places not mentioned in the document. The {\small\tt AnchorContext-QB} repeats itself and introduces ``walking'' tours. Also, it is very generous in its praise, making strong subjective claims unbefitting snippet language. This indicates some bias in the anchor context training data: Perhaps, more often than not, an author linking to another document has good things to say about it (e.g., when referring to the web document of a nice place one has visited). This merits further investigation and perhaps the inclusion of an additional filtering step based on sentiment analysis. Both models introduce errors relating to determiner usage.

\begin{table}[t]%
\small%
\centering%
\setlength{\tabcolsep}{1pt}%
\renewcommand{\arraystretch}{0.9}%
\setuldepth{Massachusetts}%
\caption{An example of a query, a relevant web document, and snippets generated by the  studied models. The query's terms are highlighted bold, reuse gray, repetitions underlined gray, and errors underlined black.}%
\label{table-examples}%
\begin{tabular}{@{}l@{}}
\toprule
\addlinespace
\parbox{\columnwidth}{{\bfseries Query}: cycling tours} \\[1ex]
\midrule
{\bfseries Web document} (excerpt; ClueWeb09 ID ``clueweb09-en0020-20-15833'')
\\[0.5ex]
\parbox{\columnwidth}{All bike new England routes are now available on one disk with over 200~cue sheets and maps from 7~years of touring in the Berkshires, the Massachusetts north shore, Cape Cod, Vermont, New Hampshire, [\,\textellipsis] This special version of our north shore {\bf cycling tours} is designed for active participants aged~55 and older through the Elderhostel travel organization exploring the England coast and historic Essex county [\,\textellipsis] (see entire document at
[\href{https://www.chatnoir.eu/cache?uuid=4fa77eab-4a40-564a-98e5-aa0397eca713&index=cw09}{ChatNoir}]) } \\
\addlinespace
\midrule
\addlinespace
\bfseries CNN-DM \\[0.5ex]
\parbox{\columnwidth}{{\color{gray}All bike new England routes are available on one disk with over {\small\tt<num>} cue sheets and maps from {\small\tt<num>} years of touring in the Berkshires, the Massachusetts north shore, Cape Cod, Vermont, New Hampshire} and {\setulcolor{lightgray}\ul{Vermont}}. {\color{gray}Rides are up to~45 miles a day, with shorter loops for all levels of {\bf cycling} ability.}} \\
\addlinespace
\midrule
\addlinespace
\bfseries Paraphraser \\[0.5ex]
\parbox{\columnwidth}{All the new England \ul{transmissions} {\color{gray}are now available on one disk with} a \ul{blanket of {\small\tt<num>}} {\color{gray}and maps from {\small\tt<num>} years of} travel in the northwest coast, \ul{a~Massachusetts north coast}, \ul{the }\textcolor{gray}{\setulcolor{black}\ul{Cape Cod}}{\color{gray}, Vermont, New Hampshire [\,\textellipsis]}} \\
\addlinespace
\midrule
\addlinespace
\bfseries DMOZ \\[0.5ex]
\parbox{\columnwidth}{Guided {\bf bike tours} in Massachusetts, {\setulcolor{lightgray}\ul{Massachusetts}}, Rhode island, New Hampshire, Connecticut, {\setulcolor{lightgray}\ul{Massachusetts}} and {\setulcolor{lightgray}\ul{Massachusetts}}. Includes information on {\bf tours}, {\setulcolor{lightgray}\ul{\bf tours}}, events, and contacts. \ul{Located in Boston.}} \\
\addlinespace
\midrule
\addlinespace
\bfseries DMOZ-QB\\[0.5ex]
\parbox{\columnwidth}{\ul{Spring }{\setulcolor{lightgray}\ul{\bf tours}}\ul{ }\ul{\bf cycling tours} offers a range of {\setulcolor{lightgray}\ul{\bf tours}} and services to individuals and groups in the north bay area. Includes information on {\setulcolor{lightgray}\ul{\bf tours}}, {\setulcolor{lightgray}\ul{\bf tours}}, events, and activities. Also includes a guide to the \ul{\small\tt <eos>}} \\
\addlinespace
\midrule
\addlinespace
\bfseries AnchorContext \\[0.5ex]
\parbox{\columnwidth}{For more information on \ul{the Deerfield river bike tour }\ul{\bf cycling}, visit the \ul{Deerfield} web site. \ul{The Deerfield is a small company} offering group tour services for organizations and individuals in the north shore area of Massachusetts, southern New Hampshire and Vermont, and \ul{the popular Worcester mountains of western Massachusetts.}} \\
\addlinespace
\midrule
\addlinespace
\bfseries AnchorContext-QB \\[0.5ex]
\parbox{\columnwidth}{\ul{Walking} and {\bf cycling tours}. \ul{The }\ul{\bf tours}\ul{ is a great place} to start and enjoy \ul{the best of the great lakes in the United States and around the world,} as well as some of \ul{the most beautiful and }{\setulcolor{lightgray}\ul{\mbox{beautiful places}}}\ul{ in }{\setulcolor{lightgray}\ul{\mbox{the world}}}.} \\
\addlinespace
\bottomrule
\end{tabular}%
\end{table}

Altogether, despite the encouraging evaluation results, our in-depth analysis of the example as well as others reveals a lot of room for improvement. All models except the mostly reusing {\small\tt CNN-DM} model introduce language or factual errors to a greater or lesser extent, the factual errors being the most important issue to tackle in future work. The baseline models {\small\tt CNN-DM} and {\small\tt Paraphraser} disqualify themselves with respect to text reuse; they are hardly abstractive. A cause for the shortcomings of the two query-biased models {\small\tt DMOZ-QB} and {\small\tt AnchorContext-QB} may be the fact that they are forced to start generating with the query, which may not be an optimal starting point, whereas query bias is important for snippet generation. 

\subsection{Model Ranking}

Snippet usefulness---the capability of a model to generate snippets that enable humans to select relevant results---is the key measure to rank abstractive snippet generation models. Our {\small\tt AnchorContext-QB} model performs best, achieving an F-score competitive to that of extractive snippets. Nevertheless, it does not achieve the crowdsourced effectiveness and fluency scores of {\small\tt CNN-DM}, which achieves the second-highest usefulness score. The latter, however, mostly reuses text from the summarized documents: There is no practical advantage in training a neural snippet generation model that is not actually abstractive, since state-of-the-art extractive snippet generators perform competitively with little development overhead.

The {\small\tt Paraphraser} and the two DMOZ-based models are ranked third to fifth in terms of usefulness, while their ranking is reversed in terms of reuse. The {\small\tt Paraphraser} and the query-biased DMOZ model have the lowest fluency among all models, while the remaining query-unbiased DMOZ model scores second to lowest in terms of usefulness. Nevertheless, the writing style of the snippets generated by the DMOZ-based models is closest to our expectation of a well-written snippet. It is conceivable, however, that by restoring the entire DMOZ directory and by retrieving archived versions of its linked pages, a substantially higher overall performance can be attained than is possible with the comparably small amount of training examples we could obtain. That size of the training data matters,  can be observed for the query-unbiased {\small\tt AnchorContext} model, which is trained on 10~million examples and achieves the best fluency in terms of perplexity and second-best fluency as per crowd judgment, while reusing the least of the original document. However, its usefulness score is lowest of all models, showing that enforcing query bias may be necessary to ensure the model does not ``hallucinate''. Thus, increasing the number of query-biased anchor context-based training examples might allow to combine the strengths of the two anchor context-based models.

\section{Conclusion}

With anchor contexts and web directory descriptions, we presented two new sources for distant supervision for the new task of abstractive snippet generation, constructing the first large-scale corpus of query-biased training examples. To effectively exploit this corpus, we propose a bidirectional generation model based on pointer-generator networks, which preserves query terms while generating fluent snippets with low text reuse from the source document. Intrinsic and extrinsic evaluations show that, dependent on what data the model uses for training, it generates abstractive snippets that can be used to reliably select relevant documents on a search results page. Nevertheless, several problems remain unsolved.

Is abstractive snippet generation worth the effort required to develop it further? That strongly depends on whether regulatory bodies will continue to limit fair use with respect to text reuse for extractive snippets, and whether generating abstractive snippets allows for improving users' search experience. Throughout the paper, we outlined several avenues for future work, and we plan on following at least some of them. We also hope that the IR~community and the natural language generation community will pick up this new challenge. It presents a fresh use case for abstractive summarization with the potential of changing an entire industry.

\begin{acks}
This work was partially supported by the German Research Foundation (DFG) within the Collaborative Research Center ``On-The-Fly Computing'' (SFB~901/3) under the project number~160364472.
\end{acks}

\begin{raggedright}



\begin{thebibliography}{39}


\ifx \showCODEN    \undefined \def \showCODEN     #1{\unskip}     \fi
\ifx \showDOI      \undefined \def \showDOI       #1{#1}\fi
\ifx \showISBNx    \undefined \def \showISBNx     #1{\unskip}     \fi
\ifx \showISBNxiii \undefined \def \showISBNxiii  #1{\unskip}     \fi
\ifx \showISSN     \undefined \def \showISSN      #1{\unskip}     \fi
\ifx \showLCCN     \undefined \def \showLCCN      #1{\unskip}     \fi
\ifx \shownote     \undefined \def \shownote      #1{#1}          \fi
\ifx \showarticletitle \undefined \def \showarticletitle #1{#1}   \fi
\ifx \showURL      \undefined \def \showURL       {\relax}        \fi
\providecommand\bibfield[2]{#2}
\providecommand\bibinfo[2]{#2}
\providecommand\natexlab[1]{#1}
\providecommand\showeprint[2][]{arXiv:#2}

\balance

\bibitem[\protect\citeauthoryear{Amidei, Piwek, and Willis}{Amidei
  et~al\mbox{.}}{2018}]%
        {amidei:2018}
\bibfield{author}{\bibinfo{person}{Jacopo Amidei}, \bibinfo{person}{Paul
  Piwek}, {and} \bibinfo{person}{Alistair Willis}.}
  \bibinfo{year}{2018}\natexlab{}.
\newblock \showarticletitle{Evaluation Methodologies in Automatic Question
  Generation 2013--2018}. In \bibinfo{booktitle}{{\em Proceedings of INLG
  2018}}. \bibinfo{pages}{307--317}.
\newblock


\bibitem[\protect\citeauthoryear{Baumel, Eyal, and Elhadad}{Baumel
  et~al\mbox{.}}{2018}]%
        {baumel:2018}
\bibfield{author}{\bibinfo{person}{Tal Baumel}, \bibinfo{person}{Matan Eyal},
  {and} \bibinfo{person}{Michael Elhadad}.} \bibinfo{year}{2018}\natexlab{}.
\newblock \showarticletitle{Query Focused Abstractive Summarization:
  Incorporating Query Relevance, Multi-Document Coverage, and Summary Length
  Constraints into Seq2seq Models}.
\newblock \bibinfo{journal}{{\em arXiv:1801.07704\/}}.
\newblock


\bibitem[\protect\citeauthoryear{Biber, Johansson, Leech, Conrad, and
  Finegan}{Biber et~al\mbox{.}}{1999}]%
        {biber:1999}
\bibfield{author}{\bibinfo{person}{Douglas Biber}, \bibinfo{person}{Stig
  Johansson}, \bibinfo{person}{Geoffrey Leech}, \bibinfo{person}{Susan Conrad},
  {and} \bibinfo{person}{Edward Finegan}.} \bibinfo{year}{1999}\natexlab{}.
\newblock \bibinfo{booktitle}{{\em Longman Grammar of Spoken and Written
  English}}.
\newblock \bibinfo{publisher}{Longman}.
\newblock


\bibitem[\protect\citeauthoryear{Brin and Page}{Brin and Page}{1998}]%
        {brin:98}
\bibfield{author}{\bibinfo{person}{Sergey Brin} {and} \bibinfo{person}{Lawrence
  Page}.} \bibinfo{year}{1998}\natexlab{}.
\newblock \showarticletitle{{The Anatomy of a Large-Scale Hypertextual Web
  Search Engine}}.
\newblock \bibinfo{journal}{{\em Computer Networks\/}}  \bibinfo{volume}{30}
  (\bibinfo{year}{1998}), \bibinfo{pages}{107--117}.
\newblock


\bibitem[\protect\citeauthoryear{Cao, Wei, Li, and Li}{Cao
  et~al\mbox{.}}{2018}]%
        {cao:2018}
\bibfield{author}{\bibinfo{person}{Ziqiang Cao}, \bibinfo{person}{Furu Wei},
  \bibinfo{person}{Wenjie Li}, {and} \bibinfo{person}{Sujian Li}.}
  \bibinfo{year}{2018}\natexlab{}.
\newblock \showarticletitle{Faithful to the Original: Fact Aware Neural
  Abstractive Summarization}. In \bibinfo{booktitle}{{\em Proceedings of AAAI
  2018}}. \bibinfo{pages}{4784--4791}.
\newblock


\bibitem[\protect\citeauthoryear{Chen, Hagen, Stein, and Potthast}{Chen
  et~al\mbox{.}}{2018a}]%
        {stein:2018k}
\bibfield{author}{\bibinfo{person}{Wei-Fan Chen}, \bibinfo{person}{Matthias
  Hagen}, \bibinfo{person}{Benno Stein}, {and} \bibinfo{person}{Martin
  Potthast}.} \bibinfo{year}{2018}\natexlab{a}.
\newblock \showarticletitle{{A User Study on Snippet Generation: Text Reuse vs.
  Paraphrases}}. In \bibinfo{booktitle}{{\em Proceedings of SIGIR 2018}}.
  \bibinfo{pages}{1033--1036}.
\newblock


\bibitem[\protect\citeauthoryear{Cormack, Smucker, and Clarke}{Cormack
  et~al\mbox{.}}{2011}]%
        {cormack:2011}
\bibfield{author}{\bibinfo{person}{Gordon~V. Cormack}, \bibinfo{person}{Mark~D.
  Smucker}, {and} \bibinfo{person}{Charles~L.A. Clarke}.}
  \bibinfo{year}{2011}\natexlab{}.
\newblock \showarticletitle{Efficient and Effective Spam Filtering and
  Re-ranking for Large Web Datasets}.
\newblock \bibinfo{journal}{{\em Information Retrieval\/}}
  \bibinfo{volume}{14}, \bibinfo{number}{5} (\bibinfo{year}{2011}),
  \bibinfo{pages}{441--465}.
\newblock


\bibitem[\protect\citeauthoryear{Dang}{Dang}{2005}]%
        {dang:2005}
\bibfield{author}{\bibinfo{person}{Hoa~Trang Dang}.}
  \bibinfo{year}{2005}\natexlab{}.
\newblock \showarticletitle{Overview of DUC 2005}. In \bibinfo{booktitle}{{\em
  Proceedings of DUC 2005}}.
\newblock


\bibitem[\protect\citeauthoryear{Dang}{Dang}{2006}]%
        {dang:2006}
\bibfield{author}{\bibinfo{person}{Hoa~Trang Dang}.}
  \bibinfo{year}{2006}\natexlab{}.
\newblock \showarticletitle{Overview of DUC 2006}. In \bibinfo{booktitle}{{\em
  Proceedings of DUC 2006}}.
\newblock


\bibitem[\protect\citeauthoryear{Dang}{Dang}{2007}]%
        {dang:2007}
\bibfield{author}{\bibinfo{person}{Hoa~Trang Dang}.}
  \bibinfo{year}{2007}\natexlab{}.
\newblock \showarticletitle{Overview of DUC 2007}. In \bibinfo{booktitle}{{\em
  Proceedings of DUC 2007}}.
\newblock


\bibitem[\protect\citeauthoryear{Devlin, Chang, Lee, and Toutanova}{Devlin
  et~al\mbox{.}}{2019}]%
        {devlin:2019}
\bibfield{author}{\bibinfo{person}{Jacob Devlin}, \bibinfo{person}{Ming-Wei
  Chang}, \bibinfo{person}{Kenton Lee}, {and} \bibinfo{person}{Kristina
  Toutanova}.} \bibinfo{year}{2019}\natexlab{}.
\newblock \showarticletitle{BERT: Pre-training of Deep Bidirectional
  Transformers for Language Understanding}. In \bibinfo{booktitle}{{\em
  Proceedings of NAACL-HLT 2019}}. \bibinfo{pages}{4171--4186}.
\newblock


\bibitem[\protect\citeauthoryear{Gatt and Krahmer}{Gatt and Krahmer}{2018}]%
        {gatt:2018}
\bibfield{author}{\bibinfo{person}{Albert Gatt} {and} \bibinfo{person}{Emiel
  Krahmer}.} \bibinfo{year}{2018}\natexlab{}.
\newblock \showarticletitle{Survey of the State of the Art in Natural Language
  Generation: Core Tasks, Applications and Evaluation}.
\newblock \bibinfo{journal}{{\em Journal of Artificial Intelligence
  Research\/}}  \bibinfo{volume}{61} (\bibinfo{year}{2018}),
  \bibinfo{pages}{65--170}.
\newblock


\bibitem[\protect\citeauthoryear{Grusky, Naaman, and Artzi}{Grusky
  et~al\mbox{.}}{2018}]%
        {grusky:2018}
\bibfield{author}{\bibinfo{person}{Max Grusky}, \bibinfo{person}{Mor Naaman},
  {and} \bibinfo{person}{Yoav Artzi}.} \bibinfo{year}{2018}\natexlab{}.
\newblock \showarticletitle{Newsroom: A Dataset of 1.3 Million Summaries with
  Diverse Extractive Strategies}. In \bibinfo{booktitle}{{\em Proceedings of
  NAACL-HLT 2018}}. \bibinfo{pages}{708--719}.
\newblock


\bibitem[\protect\citeauthoryear{Gu, Lu, Li, and Li}{Gu et~al\mbox{.}}{2016}]%
        {gu:2016}
\bibfield{author}{\bibinfo{person}{Jiatao Gu}, \bibinfo{person}{Zhengdong Lu},
  \bibinfo{person}{Hang Li}, {and} \bibinfo{person}{Victor~O.K. Li}.}
  \bibinfo{year}{2016}\natexlab{}.
\newblock \showarticletitle{Incorporating Copying Mechanism in
  Sequence-to-Sequence Learning}. In \bibinfo{booktitle}{{\em Proceedings of
  ACL 2016}}. \bibinfo{pages}{1631--1640}.
\newblock


\bibitem[\protect\citeauthoryear{Hagen, Potthast, Beyer, and Stein}{Hagen
  et~al\mbox{.}}{2012}]%
        {hagen:2012}
\bibfield{author}{\bibinfo{person}{Matthias Hagen}, \bibinfo{person}{Martin
  Potthast}, \bibinfo{person}{Anna Beyer}, {and} \bibinfo{person}{Benno
  Stein}.} \bibinfo{year}{2012}\natexlab{}.
\newblock \showarticletitle{Towards Optimum Query Segmentation: In Doubt
  Without}. In \bibinfo{booktitle}{{\em Proceedings of CIKM 2012}}.
  \bibinfo{pages}{1015--1024}.
\newblock


\bibitem[\protect\citeauthoryear{Hasselqvist, Helmertz, and
  K{\aa}geb{\"a}ck}{Hasselqvist et~al\mbox{.}}{2017}]%
        {hasselqvist:2017}
\bibfield{author}{\bibinfo{person}{Johan Hasselqvist}, \bibinfo{person}{Niklas
  Helmertz}, {and} \bibinfo{person}{Mikael K{\aa}geb{\"a}ck}.}
  \bibinfo{year}{2017}\natexlab{}.
\newblock \showarticletitle{Query-Based Abstractive Summarization using Neural
  Networks}.
\newblock \bibinfo{journal}{{\em arXiv:1712.06100\/}}.
\newblock


\bibitem[\protect\citeauthoryear{Hermann, Kocisky, Grefenstette, Espeholt, Kay,
  Suleyman, and Blunsom}{Hermann et~al\mbox{.}}{2015}]%
        {hermann:2015}
\bibfield{author}{\bibinfo{person}{Karl~Moritz Hermann}, \bibinfo{person}{Tomas
  Kocisky}, \bibinfo{person}{Edward Grefenstette}, \bibinfo{person}{Lasse
  Espeholt}, \bibinfo{person}{Will Kay}, \bibinfo{person}{Mustafa Suleyman},
  {and} \bibinfo{person}{Phil Blunsom}.} \bibinfo{year}{2015}\natexlab{}.
\newblock \showarticletitle{Teaching Machines to Read and Comprehend}. In
  \bibinfo{booktitle}{{\em Proceedings of NIPS 2015}}.
  \bibinfo{pages}{1693--1701}.
\newblock


\bibitem[\protect\citeauthoryear{Jones and Galliers}{Jones and
  Galliers}{1995}]%
        {jones:1995}
\bibfield{author}{\bibinfo{person}{Karen~Sparck Jones} {and}
  \bibinfo{person}{Julia~R Galliers}.} \bibinfo{year}{1995}\natexlab{}.
\newblock \bibinfo{booktitle}{{\em Evaluating Natural Language Processing
  Systems: An Analysis and Review}}.
\newblock \bibinfo{publisher}{Springer}.
\newblock


\bibitem[\protect\citeauthoryear{Kaisser, Hearst, and Lowe}{Kaisser
  et~al\mbox{.}}{2008}]%
        {kaisser:08}
\bibfield{author}{\bibinfo{person}{Michael Kaisser}, \bibinfo{person}{{Marti
  A.} Hearst}, {and} \bibinfo{person}{{John B.} Lowe}.}
  \bibinfo{year}{2008}\natexlab{}.
\newblock \showarticletitle{{Improving Search Results Quality by Customizing
  Summary Lengths}}. In \bibinfo{booktitle}{{\em Proceedings of ACL 2008}}.
  \bibinfo{pages}{701--709}.
\newblock


\bibitem[\protect\citeauthoryear{Kiesel, Stein, and Lucks}{Kiesel
  et~al\mbox{.}}{2017}]%
        {kiesel:2017}
\bibfield{author}{\bibinfo{person}{Johannes Kiesel}, \bibinfo{person}{Benno
  Stein}, {and} \bibinfo{person}{Stefan Lucks}.}
  \bibinfo{year}{2017}\natexlab{}.
\newblock \showarticletitle{A Large-scale Analysis of the Mnemonic Password
  Advice}. In \bibinfo{booktitle}{{\em Proceedings of NDSS 2017}}.
\newblock


\bibitem[\protect\citeauthoryear{Kim, Thomas, Sankaranarayana, Gedeon, and
  Yoon}{Kim et~al\mbox{.}}{2017}]%
        {kim:17}
\bibfield{author}{\bibinfo{person}{Jaewon Kim}, \bibinfo{person}{Paul Thomas},
  \bibinfo{person}{Ramesh Sankaranarayana}, \bibinfo{person}{Tom Gedeon}, {and}
  \bibinfo{person}{Hwan-Jin Yoon}.} \bibinfo{year}{2017}\natexlab{}.
\newblock \showarticletitle{{What Snippet Size is Needed in Mobile Web
  Search?}} In \bibinfo{booktitle}{{\em Proceedings of CHIIR 2017}}.
  \bibinfo{pages}{97--106}.
\newblock


\bibitem[\protect\citeauthoryear{Klein, Kim, Deng, Senellart, and Rush}{Klein
  et~al\mbox{.}}{2017}]%
        {klein:2017}
\bibfield{author}{\bibinfo{person}{Guillaume Klein}, \bibinfo{person}{Yoon
  Kim}, \bibinfo{person}{Yuntian Deng}, \bibinfo{person}{Jean Senellart}, {and}
  \bibinfo{person}{Alexander~M. Rush}.} \bibinfo{year}{2017}\natexlab{}.
\newblock \showarticletitle{Open{NMT}: Open-Source Toolkit for Neural Machine
  Translation}. In \bibinfo{booktitle}{{\em Proceedings of ACL 2017 (Demos)}}. \bibinfo{pages}{67--72}.
\newblock


\bibitem[\protect\citeauthoryear{Lin}{Lin}{2004}]%
        {lin:2004}
\bibfield{author}{\bibinfo{person}{Chin-Yew Lin}.}
  \bibinfo{year}{2004}\natexlab{}.
\newblock \showarticletitle{ROUGE: A package for Automatic Evaluation of
  Summaries}. In \bibinfo{booktitle}{{\em Proceedings of the ACL Workshop on
  Text Summarization Branches Out}}. \bibinfo{pages}{74--81}.
\newblock


\bibitem[\protect\citeauthoryear{Liu, Saleh, Pot, Goodrich, Sepassi, Kaiser,
  and Shazeer}{Liu et~al\mbox{.}}{2018}]%
        {liu:2018}
\bibfield{author}{\bibinfo{person}{Peter~J Liu}, \bibinfo{person}{Mohammad
  Saleh}, \bibinfo{person}{Etienne Pot}, \bibinfo{person}{Ben Goodrich},
  \bibinfo{person}{Ryan Sepassi}, \bibinfo{person}{Lukasz Kaiser}, {and}
  \bibinfo{person}{Noam Shazeer}.} \bibinfo{year}{2018}\natexlab{}.
\newblock \showarticletitle{Generating Wikipedia by Summarizing Long
  Sequences}. In \bibinfo{booktitle}{{\em Proceedings of ICLR 2018 (Posters)}}.
\newblock


\bibitem[\protect\citeauthoryear{Luhn}{Luhn}{1958}]%
        {luhn:1958}
\bibfield{author}{\bibinfo{person}{{Hans Peter} Luhn}.}
  \bibinfo{year}{1958}\natexlab{}.
\newblock \showarticletitle{{The Automatic Creation of Literature Abstracts}}.
\newblock \bibinfo{journal}{{\em {IBM} Journal of Research and Development\/}}
  \bibinfo{volume}{2}, \bibinfo{number}{2} (\bibinfo{year}{1958}),
  \bibinfo{pages}{159--165}.
\newblock


\bibitem[\protect\citeauthoryear{Maxwell, Azzopardi, and Moshfeghi}{Maxwell
  et~al\mbox{.}}{2017}]%
        {maxwell:17}
\bibfield{author}{\bibinfo{person}{David Maxwell}, \bibinfo{person}{Leif
  Azzopardi}, {and} \bibinfo{person}{Yashar Moshfeghi}.}
  \bibinfo{year}{2017}\natexlab{}.
\newblock \showarticletitle{{A Study of Snippet Length and Informativeness:
  Behaviour, Performance and User Experience}}. In \bibinfo{booktitle}{{\em
  Proceedings of SIGIR 2017}}. \bibinfo{pages}{135--144}.
\newblock


\bibitem[\protect\citeauthoryear{Nema, Khapra, Laha, and Ravindran}{Nema
  et~al\mbox{.}}{2017}]%
        {nema:2017}
\bibfield{author}{\bibinfo{person}{Preksha Nema}, \bibinfo{person}{Mitesh~M
  Khapra}, \bibinfo{person}{Anirban Laha}, {and} \bibinfo{person}{Balaraman
  Ravindran}.} \bibinfo{year}{2017}\natexlab{}.
\newblock \showarticletitle{Diversity Driven Attention Model for Query-based
  Abstractive Summarization}. In \bibinfo{booktitle}{{\em Proceedings of ACL
  2017}}. \bibinfo{pages}{1063--1072}.
\newblock


\bibitem[\protect\citeauthoryear{Potthast, Chen, Hagen, and Stein}{Potthast
  et~al\mbox{.}}{2018}]%
        {stein:2018b}
\bibfield{author}{\bibinfo{person}{Martin Potthast}, \bibinfo{person}{Wei-Fan
  Chen}, \bibinfo{person}{Matthias Hagen}, {and} \bibinfo{person}{Benno
  Stein}.} \bibinfo{year}{2018}\natexlab{}.
\newblock \showarticletitle{{A Plan for Ancillary Copyright: Original
  Snippets}}. In \bibinfo{booktitle}{{\em Proceedings of the NewsIR workshop at ECIR~2018}}.
  \bibinfo{pages}{3--5}.
\newblock


\bibitem[\protect\citeauthoryear{Rao and Zhu}{Rao and Zhu}{2016}]%
        {rao:2016}
\bibfield{author}{\bibinfo{person}{BiChen Rao} {and} \bibinfo{person}{Erkang
  Zhu}.} \bibinfo{year}{2016}\natexlab{}.
\newblock \showarticletitle{Searching Web Data using MinHash LSH}. In
  \bibinfo{booktitle}{{\em Proceedings of {SIGMOD} 2016}}.
  \bibinfo{pages}{2257--2258}.
\newblock


\bibitem[\protect\citeauthoryear{Rush, Chopra, and Weston}{Rush
  et~al\mbox{.}}{2015}]%
        {rush:2015}
\bibfield{author}{\bibinfo{person}{{Alexander M.} Rush}, \bibinfo{person}{Sumit
  Chopra}, {and} \bibinfo{person}{Jason Weston}.}
  \bibinfo{year}{2015}\natexlab{}.
\newblock \showarticletitle{{A Neural Attention Model for Abstractive Sentence
  Summarization}}. In \bibinfo{booktitle}{{\em Proceedings of EMNLP 2015}}.
  \bibinfo{pages}{379--389}
\newblock


\bibitem[\protect\citeauthoryear{See, Liu, and Manning}{See
  et~al\mbox{.}}{2017}]%
        {see:2017}
\bibfield{author}{\bibinfo{person}{Abigail See}, \bibinfo{person}{{Peter J.}
  Liu}, {and} \bibinfo{person}{{Christopher D.} Manning}.}
  \bibinfo{year}{2017}\natexlab{}.
\newblock \showarticletitle{{Get to the Point: Summarization with
  Pointer-Generator Networks}}. In \bibinfo{booktitle}{{\em Proceedings of ACL
  2017}}. \bibinfo{pages}{1073--1083}
\newblock


\bibitem[\protect\citeauthoryear{Syed, V{\"o}lske, Lipka, Stein, Sch{\"u}tze,
  and Potthast}{Syed et~al\mbox{.}}{2019}]%
        {stein:2019x}
\bibfield{author}{\bibinfo{person}{Shahbaz Syed}, \bibinfo{person}{Michael
  V{\"o}lske}, \bibinfo{person}{Nedim Lipka}, \bibinfo{person}{Benno Stein},
  \bibinfo{person}{Hinrich Sch{\"u}tze}, {and} \bibinfo{person}{Martin
  Potthast}.} \bibinfo{year}{2019}\natexlab{}.
\newblock \showarticletitle{{Towards Summarization for Social Media - Results
  of the TL;DR Challenge}}. In \bibinfo{booktitle}{{\em Proceedings of INLG~2019}}. \bibinfo{pages}{523--528}
\newblock


\bibitem[\protect\citeauthoryear{Syed, V{\"o}lske, Potthast, Lipka, Stein, and
  Sch{\"u}tze}{Syed et~al\mbox{.}}{2018}]%
        {stein:2018za}
\bibfield{author}{\bibinfo{person}{Shahbaz Syed}, \bibinfo{person}{Michael
  V{\"o}lske}, \bibinfo{person}{Martin Potthast}, \bibinfo{person}{Nedim
  Lipka}, \bibinfo{person}{Benno Stein}, {and} \bibinfo{person}{Hinrich
  Sch{\"u}tze}.} \bibinfo{year}{2018}\natexlab{}.
\newblock \showarticletitle{{Task Proposal: The TL;DR Challenge}}. In
  \bibinfo{booktitle}{{\em Proceedings of INLG 2018}}.
  \bibinfo{pages}{318--321}.
\newblock


\bibitem[\protect\citeauthoryear{Tombros and Sanderson}{Tombros and
  Sanderson}{1998}]%
        {tombros:1998}
\bibfield{author}{\bibinfo{person}{Anastasios Tombros} {and}
  \bibinfo{person}{Mark Sanderson}.} \bibinfo{year}{1998}\natexlab{}.
\newblock \showarticletitle{{Advantages of Query Biased Summaries in
  Information Retrieval}}. In \bibinfo{booktitle}{{\em Proceedings of {SIGIR}
  1998}}. \bibinfo{pages}{2--10}.
\newblock


\bibitem[\protect\citeauthoryear{Toutanova, Klein, Manning, and
  Singer}{Toutanova et~al\mbox{.}}{2003}]%
        {toutanova:2003}
\bibfield{author}{\bibinfo{person}{Kristina Toutanova}, \bibinfo{person}{Dan
  Klein}, \bibinfo{person}{Christopher~D. Manning}, {and}
  \bibinfo{person}{Yoram Singer}.} \bibinfo{year}{2003}\natexlab{}.
\newblock \showarticletitle{Feature-Rich Part-of-Speech Tagging with a Cyclic
  Dependency Network}. In \bibinfo{booktitle}{{\em Proceedings of NAACL-HLT
  2003}}. \bibinfo{pages}{252--259}.
\newblock


\bibitem[\protect\citeauthoryear{White, Ruthven, and Jose}{White
  et~al\mbox{.}}{2002a}]%
        {white:2002a}
\bibfield{author}{\bibinfo{person}{Ryen White}, \bibinfo{person}{Ian Ruthven},
  {and} \bibinfo{person}{{Joemon M.} Jose}.} \bibinfo{year}{2002}\natexlab{a}.
\newblock \showarticletitle{{Finding Relevant Documents using Top Ranking
  Sentences: An Evaluation of Two Alternative Schemes}}. In
  \bibinfo{booktitle}{{\em Proceedings of {SIGIR} 2002}}.
  \bibinfo{pages}{57--64}.
\newblock


\bibitem[\protect\citeauthoryear{White, Ruthven, and Jose}{White
  et~al\mbox{.}}{2002b}]%
        {white:2002b}
\bibfield{author}{\bibinfo{person}{Ryen White}, \bibinfo{person}{Ian Ruthven},
  {and} \bibinfo{person}{{Joemon M.} Jose}.} \bibinfo{year}{2002}\natexlab{b}.
\newblock \showarticletitle{{The Use of Implicit Evidence for Relevance
  Feedback in Web Retrieval}}. In \bibinfo{booktitle}{{\em Proceedings of ECIR
  2002}}. \bibinfo{pages}{93--109}.
\newblock


\bibitem[\protect\citeauthoryear{Wieting, Mallinson, and Gimpel}{Wieting
  et~al\mbox{.}}{2017}]%
        {wieting:2017}
\bibfield{author}{\bibinfo{person}{John Wieting}, \bibinfo{person}{Jonathan
  Mallinson}, {and} \bibinfo{person}{Kevin Gimpel}.}
  \bibinfo{year}{2017}\natexlab{}.
\newblock \showarticletitle{Learning Paraphrastic Sentence Embeddings from
  Back-Translated Bitext}. In \bibinfo{booktitle}{{\em Proceedings of EMNLP
  2017}}. \bibinfo{pages}{274--285}
\newblock


\end{thebibliography}
\end{raggedright}

\end{document}